\documentclass[%
 aip,
rsi,%
 amsmath,amssymb,
preprint,%
]{revtex4-1}

\usepackage{graphicx}% Include figure files
\usepackage{bm}% bold math
\begin{document}

\preprint{AIP/123-QED}

\title[Optical Neural Networks: The 3D connection]{Optical Neural Networks: The 3D connection}

\author{Niyazi Ulas Dinc}
\affiliation{Optics Laboratory, \'{E}cole Polytechnique F\'{e}d\'{e}rale de Lausanne, Lausanne, Switzerland.}%

\author{Demetri Psaltis}
\affiliation{Optics Laboratory, \'{E}cole Polytechnique F\'{e}d\'{e}rale de Lausanne, Lausanne, Switzerland.}%

\author{Daniel Brunner}
\email{daniel.brunnerfemto-st.fr}
\affiliation{D\'{e}partement d'Optique P. M. Duffieux, Institut FEMTO-ST,  Universit\'e Bourgogne-Franche-Comt\'e CNRS UMR 6174, Besan\c{c}on, France.}%

\date{\today}% It is always \today, today,

\begin{abstract}

We motivate a new canonical strategy for integrating photonic neural networks (NN) by leveraging 3D printing. Our believe is that a NN’s parallel and dense connectivity is not scalable without 3D integration. 3D additive fabrication complemented with photonic signal transduction can dramatically augment the current capabilities of 2D CMOS and integrated photonics. Here we review some of our recent advances made towards such a breakthrough architecture.

\end{abstract}

\maketitle

\section{\label{sec:Intro}Introduction}

%%%%%%%%%%% Fig. 1 %%%%%%%%%%%%%%%%%%%%%%%%%%%%%%%%%%%%
\begin{figure}[t]
	\center{\includegraphics[width=1\linewidth]{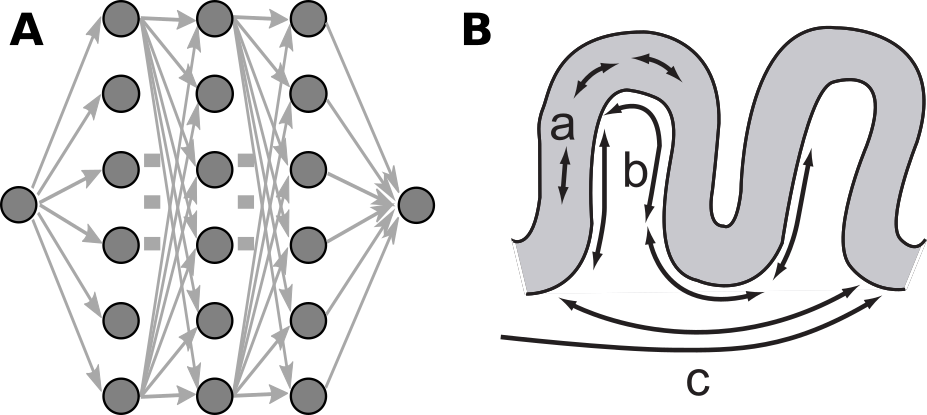}}
	\caption[]{(A) In a Neural Network (NN) typically millions of connections link simple nonlinear neurons which are arranged in layers. (B) In the brain short, medium and long range (a, b, c, respectively) neural connections are established in the volume of white and grey matter. Adapted from Schüz, et al., Encyclopedia of Neuroscience 2009.}
	\label{fig:Figure1}
\end{figure}
%%%%%%%%%%%%%%%%%%%%%%%%%%%%%%%%%%%%%%%%%%%%%%%%%%%%%%%

Several decades passed between the introduction and the large-scale exploration of neural networks (NN). Since the proposal of simple NNs in 1943 \cite{McCulloch1943}, the field has gone through multiple cycles of euphoria and challenges until reaching today’s large-scale interest and exploitation \cite{LeCun2015}. Readily available high-performance computing systems now allow emulating powerful (deep) NN architectures whose connections are optimized based on computationally expensive learning concepts such as gradient back-propagation. As a consequence, NNs currently excel on previously unseen scales, but at the same time the constraints of today’s CMOS-based computing threatens to limit the reach of this revolution.
 
As illustrated by their name, the initial objective of NN, cf. Fig. \ref{fig:Figure1}(A), was providing a ‘logical calculus of the ideas immanent in nervous activity’ \cite{McCulloch1943}, and as such their composition mirrors a most rudimentary aspect of the mammalian neo-cortex: nodes are densely linked into a network with connections much like synapses, dendrites and axons connecting biological neurons. However, this is only possible in the context of a global structural property of the neocortex in which neurons, and even more so connections, are distributed across a 3D volume, cf. Fig. \ref{fig:Figure1}(B). The majority of cortical neurons are arranged in planes located inside the grey matter that wraps around the brain, and stacks of neurons form short-range connections (labelled a in Fig. \ref{fig:Figure1}(B)) which travers the grey matter’s volume. Crucially, grey matter encloses white matter, and inside this volume the brain’s long-range (labelled b and c in Fig. \ref{fig:Figure1}(B)) connections are located. 3D connections are therefore a canonical feature of brain architecture. The scale and connectivity of the human brain’s network would otherwise simply not fit inside the human skull. The brain therefore provides a very good primer for exploiting 3D circuit topology. Even though the 3D topology of brains emerged from evolutionary development, science and engineering can deliberately combine advantageous strategies and concepts. Combining the 3D network topology of biological brains with photonic signal transduction is a highly appealing strategy for next generation NN computing.
In this paper, we elaborate the potential of 3D printing technology for integrated photonic NN chips. Such additive fabrication enables true 3D integration and naturally complements the mostly 2D lithography that struggles to implement parallel NN connections with a scalable strategy. Photonics offers fundamental energy, speed and latency advantages when establishing the communication between NN neurons along the staggering amount of network connections. 3D printing is a potential path for 3D integration of optically interconnected Si or other electro-optic chips.

\section{\label{sec:CanonicalNetwork}Canonical 3D photonic neural network architecture}

%%%%%%%%%%% Fig. 2 %%%%%%%%%%%%%%%%%%%%%%%%%%%%%%%%%%%%
\begin{figure}[t]
	\center{\includegraphics[width=1\linewidth]{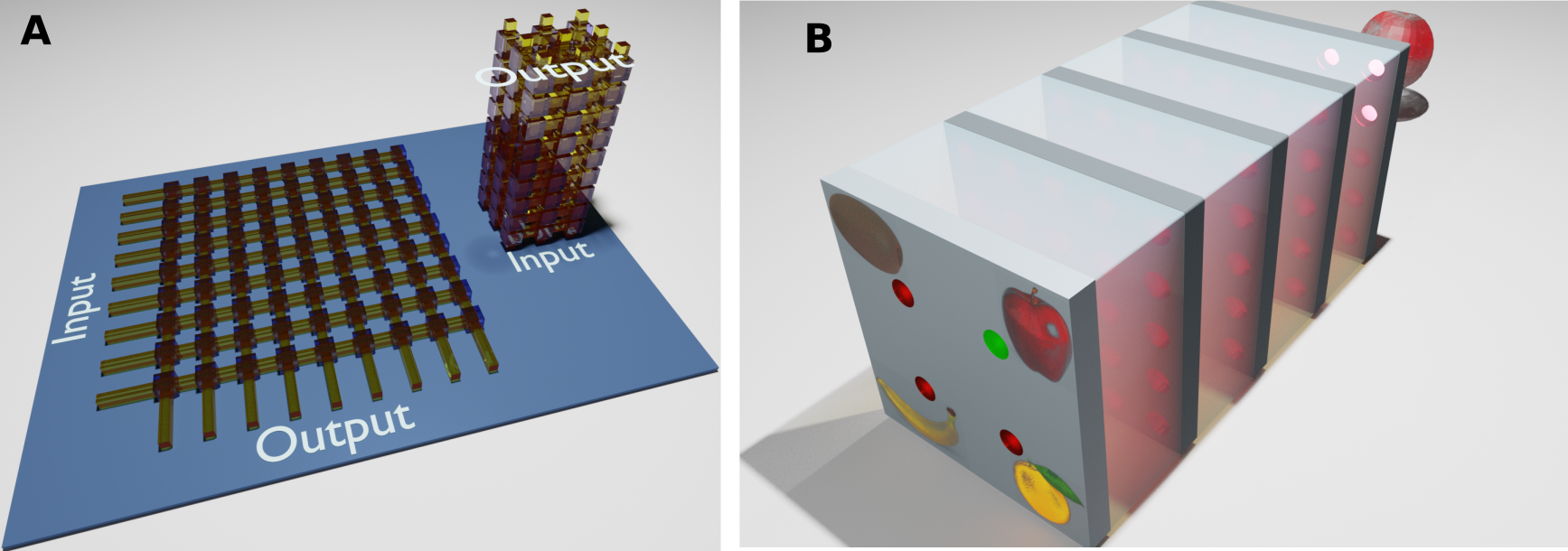}}
	\caption[]{(A) Realizing the connections in 2D interconnects is not scalable, and 3D integration is essential for parallel NN integration. (B) In our canonical 2D/3D photonic NN neurons are arranged in 2D while connections are established in the 3D volume between layers of neurons, where the NN correctly identifies an apple.}
	\label{fig:Figure2}
\end{figure}
%%%%%%%%%%%%%%%%%%%%%%%%%%%%%%%%%%%%%%%%%%%%%%%%%%%%%%%

Physically realizing dense connections for the large number of neurons (typically >1000 units) contained in each NN layer results in a formidable challenge. A parallel NN processor needs to provide a dedicated physical link for each connection, which is difficult since the amount of possible connections scales quadratically with the number of neurons. A connection’s defining property is its strength, and its physical implementation for example by memristors, micro-rings or holographic memory always occupies some basic unit of area i.e. volume. Integration in 2D results in a quadratic scaling of the circuit’s area with a network’s size \cite{Moughames2020}, cf. Fig. \ref{fig:Figure2}(A). In a 3D implementation weights can be stacked, for example, in planes, and for the simplest organization \cite{Moughames2020}, both, the number of required planes and memory-elements per plane scale linearly with the number of neurons. This mitigates the size-scalability roadblock and 3D routing may well be a fundamental prerequisite for scalable and parallel NN chips. Realizing such 3D circuits electronically is challenging due to the capacitive coupling and the associated energy dissipation when sending information along signalling wires.

In order to overcome these challenges we investigated a new, canonical photonic NN architecture where neurons in the form of nonlinear components are arranged in 2D sheets, while connections are integrated in 3D printed photonic circuits, cf. Fig. \ref{fig:Figure2}(B). We do not constrain the nature of photonic neurons or the 3D routing strategy. All-optical as well as electro-optical components acting as neurons are possible, and the 3D photonic interconnect can be realized by refractive index modifications in a 3D medium, multiple stacks of diffractive-optics planes \cite{Dinc2020} as well as complex 3D circuitry of photonic waveguides \cite{Moughames2020}.

\section{\label{sec:TPP}3D nano-printing technology}

%%%%%%%%%%% Fig. 3 %%%%%%%%%%%%%%%%%%%%%%%%%%%%%%%%%%%%
\begin{figure}[t]
	\center{\includegraphics[width=1\linewidth]{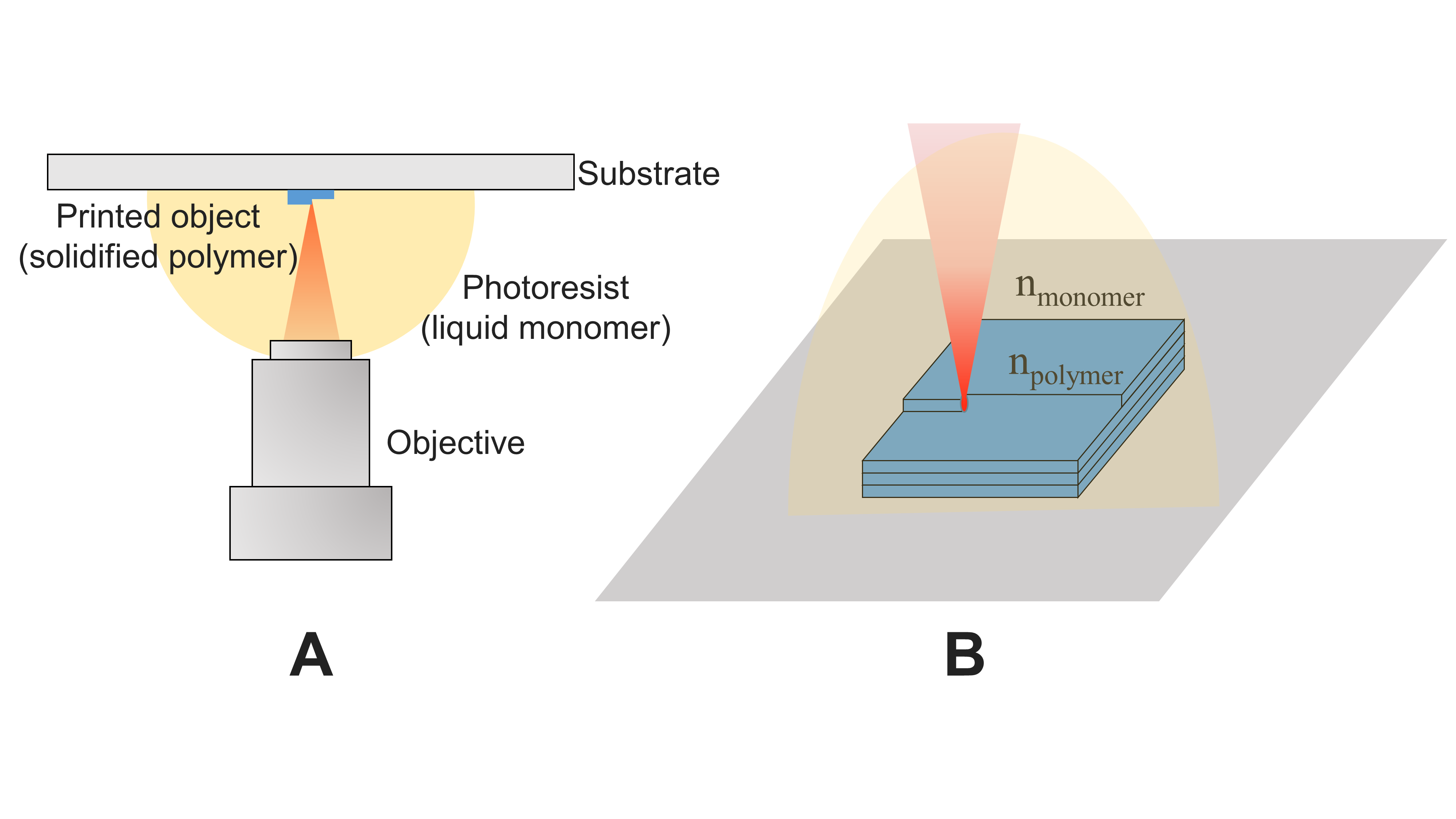}}
	\caption[]{(A) 3D printing scheme with the objective focusing the femtosecond laser pulse into the photoresist. (B) Layer-by-layer printing.}
	\label{fig:Figure3}
\end{figure}
%%%%%%%%%%%%%%%%%%%%%%%%%%%%%%%%%%%%%%%%%%%%%%%%%%%%%%% 

Additive manufacturing (AM) has been a popular method for prototyping ever since it was developed in the 1980s as it does not require special tooling or molds. However, its true advantage over most conventional manufacturing methods is AM’s ability to produce 3D parts of great complexity, which is unfeasible or even impossible with subtractive or 2D lithographic methods. Among various AM techniques, two-photon polymerization (TPP) is of special interest since it provides sub-micron feature sizes in materials that are transparent in the optical domain with refractive index values close to those of glass. TPP utilizes femtosecond lasers to expose and polymerize photoresists. The two-photon process is of significance as it enables feature sizes below the Abbe diffraction limit thanks to the polymerization’s quadratic dependence on exposure intensity. One-photon processes in turn yield larger polymerized voxels due to a linear dependence of polymerization on exposure intensity. Control of the light intensity threshold for polymerization and quenching effects further contribute to sub-diffraction resolution. TPP exposure-dose can be controlled through scanning speed and laser intensity, which provides control over the degree of polymerization of the photoresist and hence over the local refractive index. This enables the possibility of printing graded-index (GRIN) elements \cite{Zukauskas2015}. 3D direct-laser writing systems offer robust, commercial TPP setups where complex optical elements can be printed (c.f. Fig. \ref{fig:Figure3}) at different resolutions by selecting among different resin-objective pairs. In subsequent sections, we present different optical elements that were fabricated by a Nanoscribe 3D printer.

For the concepts presented in this paper, the most important feature of AM/TPP is the ability to access independently each voxel in the fabrication volume, which enables holographic as well as wave-guide based photonic connections. From the holography point of view it is key to go beyond 1/M2, which is the efficiency relation where M is the number of multiplexed holograms \cite{Psaltis1990}. This fundamental limitation holds for any optical holographic material where recording is accomplished by means of multiple optical exposures \cite{Barbastathis2000} due to the superposition of multiple holograms following a recording sequence that is designed to use the dynamic range of the index modulation equally. Crucially, efficiency could be improved to 1/M if the hologram were constructed voxel-by-voxel or in a multilayered fashion. TPP makes it practical to adopt both options. In addition, the ability to access each point in the volume enables the fabrication of complex 3D-routed waveguides that define the optical signal’s path in 3D, reminiscent of the dendrites and axons in the brain.

\section{\label{sec:3DDiscreteWG}3D discrete-waveguide interconnects}

%%%%%%%%%%% Fig. 4 %%%%%%%%%%%%%%%%%%%%%%%%%%%%%%%%%%%%
\begin{figure}[t]
	\center{\includegraphics[width=1\linewidth]{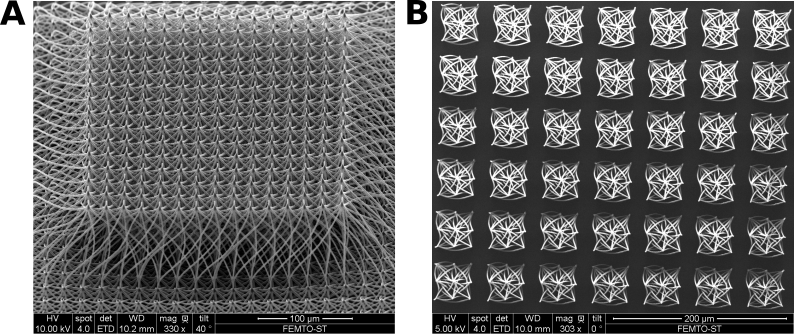}}
	\caption[]{SEM micrographs of 3D printed waveguides \cite{Moughames2020} realizing parallel interconnects with high connectivity (A) and according to Haar filters (B).}
	\label{fig:Figure4}
\end{figure}
%%%%%%%%%%%%%%%%%%%%%%%%%%%%%%%%%%%%%%%%%%%%%%%%%%%%%%% 

As previously introduced, connections between biological neurons are made by dedicated ‘wires’ formed by axons connected to dendrites via synapses, and the photonic equivalent of such spatially discrete links is the optical waveguide. An optical waveguide utilizes the principle of total internal reflection, where a medium with a higher refractive index is surrounded by a medium with a lower refractive index. Recently Moughames et al. \cite{Moughames2020} 3D printed such optical waveguides using a Nanoscribe 3D printer and connections in the form of optical splitters realized the dense connectivity between neurons. 

Different connection topologies were demonstrated. Arranging 1 to 81 splitters in an 15x15 input waveguide array, c.f. Fig. \ref{fig:Figure4}(A), demonstrated a 3D printed dense interconnect for 225 neurons in an area of only 300x300 µm2. Inspired by convolutional NNs, the same authors realized Boolean Haar filters arranged in a 7x7 array, see Fig. \ref{fig:Figure4}(B). Such arrays can filter images containing 21x21 pixels in parallel, which in principle is sufficient for realizing a convolutional layer applied to the MNIST handwritten digit dataset. Most importantly the area of both 3D interconnects scales linearly with the number of inputs.

\section{\label{sec:GRIN}Gradient index continuous interconnects}

%%%%%%%%%%% Fig. 5 %%%%%%%%%%%%%%%%%%%%%%%%%%%%%%%%%%%%
\begin{figure}[t]
	\center{\includegraphics[width=1\linewidth]{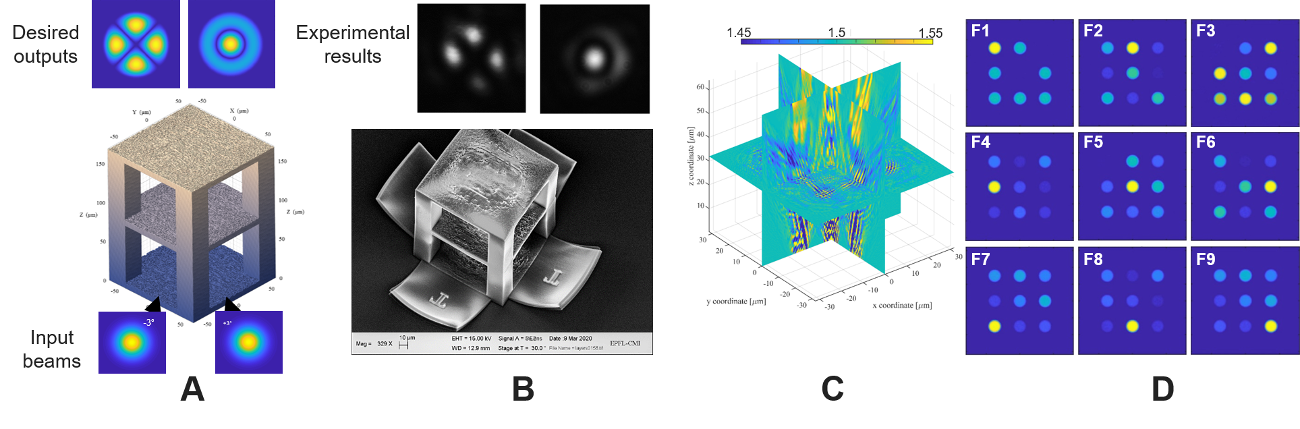}}
	\caption[]{(A) 3D rendering of the OVE in \cite{Dinc2020} with the ideal input and output pairs; (B) SEM image of the printed structure and the corresponding experimental results. (C) XY, YZ and XZ cut planes of a GRIN optimized for Haar filtering. The colorbar shows RI variation. (D) Corresponding output fields obtained by simulating the propagation of inputs through the optimized GRIN volume. All field plots have a window size of 32x32 μm2 and color code shows the normalized amplitude for each.}
	\label{fig:Figure5}
\end{figure}
%%%%%%%%%%%%%%%%%%%%%%%%%%%%%%%%%%%%%%%%%%%%%%%%%%%%%%% 

Multilayered diffractive optical elements, c.f. Fig. \ref{fig:Figure5}, can also perform interconnection tasks utilizing the 3D via optical volume elements (OVEs). OVEs can be designed by utilizing a nonlinear optimization scheme, learning tomography (LT), which calculates the topography of either multilayered or GRIN volume elements to approximate desired mappings. Figure 5(A,B) shows an demonstration by Dinc et al., which acts as an angular multiplexer (lantern) that maps plane waves with different incidence angles to linearly polarized multimode fiber modes \cite{Dinc2020}. It provides an interconnect between single mode fibers stacked with different angles and a multimode fiber to map each single mode fiber input/output to a specific mode of multimode fiber, hence performs mode-division multiplexing. Another example of LT computed OVEs realizing Haar filters such as demonstrated in \cite{Moughames2020} are shown in Fig. \ref{fig:Figure5}(C,D).

\section{\label{sec:Possibilities}Possibilities for photonic neurons}

The function of a NN neuron is the summation of its inputs followed by a nonlinear transformation. Summation of the individual fields impinging on a neuron can be realized in photonics by the superposition of optical fields. Unfortunately, nonlinearity is since many years the Achilles-heel of photonics compared to electronics. However, modern photonic devices have significantly lowered the energy consumption which can now be below 100 fJ per nonlinear transformation \cite{Heuser2020}. Many standard nonlinear photonic components have potentially high modulation bandwidths, fast response times and can directly be interfaced with fully parallel as well as dense 3D photonic interconnect. Photonic neurons combined with our 2D/3D canonical NN architecture therefore offer new concepts for addressing the long-standing challenges of parallelism and connection density for high-speed NN computers.

In order to make most efficient use of the footprint and circuit volume, photonic neurons need to be arranged in a 2D array. Furthermore, neurons that accept multi-mode fields as their input could potentially be beneficial as this relaxes design constraints and allows for high-density integration of 3D photonic waveguides without a cladding. Finally, any optical transformation is associated with losses and the 3D photonic interconnect is no exception; neurons including optical amplification would mitigate such losses. At this stage, we can imagine all-optical, electro-optical as well as plasmonic neurons, and the most promising concept will certainly have to strike a balance between speed, efficiency, flexibility and potentially amplification.

\section{\label{sec:Outlook}Outlook}

The viability of integrating photonic circuits suited for NN interconnects in 3D has recently been demonstrated in principle \cite{Moughames2020,Dinc2020}. Ultimately, scalability is key for computing hardware, which implies that stacking 2D neurons and 3D interconnects into deep photonic NNs requires optical losses to be counterbalanced by amplification without resulting in an unsustainable thermal energy deposition inside the integrated photonic circuit.
However, the computational power of a NN relies on more than simply establishing specific connections in parallel. The nonlinearity of its neurons is a fundamental requirement for solving complex tasks, and here significant room for improvement exists. Another defining feature of NN is the optimization of their connections during training. New, ideally in-situ optimization strategies are in urgent demand. In combination with plasticity such as non-volatile memristive effects, these concepts would significantly reduce the complexity of potential auxiliary support circuits as well as of the 3D interconnect itself.

\nocite{*}

\bibliography{references}% Produces the bibliography via BibTeX.

\end{document}